\documentclass[12pt,preprint]{aastex}

\newcommand{\rxte}{{\it RXTE~}}
\newcommand{\ulysses}{{\it Ulysses~}}
\newcommand{\chandra}{{\it Chandra~}}

\newcommand{\sax}{{\it BeppoSAX~}}

\newcommand{\sgr}{{SGR~$1900+14$~}}

\begin{document}

\title{Multi-wavelength observations of the Soft Gamma Repeater
SGR~$1900+14$
during its April 2001 activation}

\author{C. Kouveliotou\altaffilmark{1,2}, A. Tennant\altaffilmark{2},
P. M. Woods\altaffilmark{1}, M. C. Weisskopf\altaffilmark{2}, K.
Hurley\altaffilmark{3}, R. P. Fender\altaffilmark{4}, S.T.
Garrington\altaffilmark{5}, S. K. Patel\altaffilmark{1}, and 
E. G\"o\u{g}\"u\c{s}\altaffilmark{6}}

\authoremail{chryssa.kouveliotou@msfc.nasa.gov}

\altaffiltext{1} {Universities Space Research Association, NSSTC,
SD-50, 320 Sparkman Drive, Huntsville, AL 35805, USA}
\altaffiltext{2} {NASA/Marshall Space Flight Center, NSSTC, SD-50, 320
Sparkman Drive, Huntsville, AL 35805, USA}
\altaffiltext{3} {University of California, Berkeley, Space Sciences
Laboratory, Berkeley, CA 94720-7450, USA}
\altaffiltext{4} {Astronomical Institute ``Anton Pannekoek''and Center
for High Energy Astrophysics, University of Amsterdam, Kruislaan 403,
1098 SJ Amsterdam, The Netherlands}
\altaffiltext{5} {University of Manchester, Nuffield Radio Astronomy
Laboratories, Jodrell bank, Cheshire, SK11 9DL, UK}
\altaffiltext{6} {Department of Physics, University of Alabama in
Huntsville, Huntsville, AL 35899, USA}

\begin{abstract}

The soft-gamma repeater \sgr became active on 18 April 2001 after
about two years of quiescence; it had remained at a very low state of
activity since the fall of 1998, when it exhibited extraordinary
flaring. We have observed the source in the gamma and X rays with
\ulysses and \chandra, and in the radio with MERLIN. We report here
the confirmation of a two component X-ray spectrum (power law $+$
blackbody), indicating  emission from the neutron star surface. We
have determined that there is a dust halo surrounding the source that
extends up to $\gtrsim100^{\prime\prime}$ from the center of \sgr,
due to scattering in the interstellar medium. 

\end{abstract}


\section{Introduction}

Soft-gamma repeaters (SGRs) are neutron stars that can be found in a
quiescent or in an active state. To date, only four SGRs are
confirmed; a fifth SGR source may have been detected twice
(\cite{cline00}) but its existence is not yet firmly established. It
was recently found (\cite{cknature,hurley99b}) that SGRs are pulsars
with periods ranging within $5<P<8$ s, and that these pulsations show
rapid spindown of the order of $\sim10^{-10}$ s/s. As argued by
Kouveliotou et al. (1998, 1999) this spindown is due to magnetic dipole
radiation (and a relativistic particle wind); the corresponding
dipolar component of their magnetic fields exceeds $10^{14}$ G, thus
establishing SGRs as `magnetars', objects initially conjectured in the
early 1990s (\cite{duncan92,thompson95}). 

What distinguishes these transient soft $\gamma-$ray bursters from the
`classic' $\gamma-$ray burst sources is the recurrence of their
activity and the much softer spectra of their bursts. When active,
SGRs emit bunches of up to hundreds of very short, low-energy bursts.
The bursts vary in duration and temporal structure from simple, single
pulses lasting less than 10 ms, to longer, and more complex events of
$\sim100$ ms; occasionally highly complex events have been recorded,
that comprise over 40 very short subpulses each lasting tens of ms
(\cite{gogus01}). 

Very rarely, SGRs emit flares, extremely energetic events (fluences
are typically $\sim10^{44}$ ergs) that are much longer (of the order
of several minutes) and exhibit a very hard initial spike ($\sim$ 100
ms) and a long soft tail, strongly modulated by the spin of the
neutron star. Only two such flares have been detected so far: the 5
March 1979 event from SGR~0526$-$66 (\cite{mazets}) and the 27 August
1998 event from \sgr (\cite{hurley99a,feroci01a}). It is becoming
evident, however, that there is a continuum of burst intensities. This
is especially clear for \sgr (where we have the largest sample) but is
also observed in SGR~1627$-$41 (\cite{mazets99,woods99b}). The
intermediate size events are roughly $10-100$ times less energetic
than SGR flares, but have a higher energy content ($\sim 10-50$ times
higher) and are longer than the average SGR burst and relatively
uncommon. 

We report here on gamma and X-ray observations obtained with \ulysses,
\chandra and on radio observations obtained with MERLIN, soon after the
recent reactivation of \sgr, heralded by an intermediate size burst
(recorded with the Italian-Dutch satellite \sax) on 18, April 2001
(\cite{1041}). Sections 1, 2 and 3 focus on the observations: upper
limits in radio wavelengths and spectral and timing results in X-rays.
We have searched for evidence of extended emission associated with a
dust halo; the source line of sight has a very large $N_{\rm H}$ and
such a halo is more or less expected. Our results are discussed in
Section 4, where we explore the implications of the current flux
level, spectra, timing and halo results for the magnetar model of
SGRs.

\section{\ulysses and MERLIN Observations}

On 18 April 2001 \sax was triggered by an intense X-ray burst  from
\sgr (\cite{1041,1043}). The same event was also recorded with {\it
Ulysses}. Triangulation confirmed that its arrival direction was
consistent with the position of \sgr. However, the duration and time
history of this burst were quite different from typical SGR bursts.
The event lasted $\sim$40 s, and its time history was modulated with
the 5.16 s period of \sgr (\cite{hurley99b}). Because the \ulysses
spacecraft was experiencing an intense solar proton flux, the GRB
experiment did not trigger on this burst; it was therefore recorded
with 0.5 s resolution, and no spectral data are available for it.
Nevertheless, we can estimate the peak flux and fluence by assuming
that the spectrum may be described by an optically thin thermal
bremsstrahlung function with $kT\sim$30 keV (\cite{iauc7611}).  The
$25-100$ keV fluence is then $\sim2.6\times10^{-4}$ erg/cm$^2$, and
the peak flux (integrated over 0.5 s) is $\sim1.7\times10^{-5}$
erg/cm$^2$ s.  These values are several orders of magnitude greater
than those corresponding to the more typical short SGR bursts but
roughly 25 and 200 times smaller in fluence and peak flux,
respectively, than the giant flare of 27 August 1998
(\cite{hurley99a,feroci01a}). The energetics and the unusual time
history of this burst qualify it as an ``intermediate event.''  

We obtained observations of \sgr with MERLIN starting $\sim$1.6 days
after this reactivation.  MERLIN is an array of 6 radio telescopes
with maximum and minimum baselines of 217 and 10 km connected with
30-MHz bandwidth microwave links to a real-time correlator. \sgr  was
observed by MERLIN at 4994 MHz between 23:10 and 11:30 UT on 19/20
April 2001. A 20 arcsec square field was imaged centered on the given
co-ordinates of \sgr, with an effective resolution of 75 mas.  There
were no obvious detected sources in this field. The map co-ordinates
are aligned to within $\sim$20 mas of the International Celestial
Reference Frame via the phase referencing technique. Given the
extremely well known location of the source, (\cite{frail99}), a
reasonable (3$\sigma$) upper limit to the 5 GHz radio flux density of
\sgr is 0.45 mJy.

\section{\chandra Observations}

\chandra observed \sgr on 22 and 30 April, 2001, starting at
04:39:18 UT and 23:09:50 UT, respectively. The two sets of 20 ks
observations were both taken with ACIS-S3 in the Continuous Clocking
(CC) mode. We detect a single burst of 14 photons in the first data
set, which appears to be real. Due to the small number of detected
photons, however, we will not discuss the event any further and we
will restrict ourselves to pulse timing, spectral and spatial analysis
of the persistent source as described below.

\subsection{Timing Analysis} 

In CC mode data, the charge packet generated by a photon is clocked
out of the detector at the rate of 2.85 msec per row.  Thus for a
photon that interacts near the center of a CCD, it will take $\sim$4.3
s until the time the charge is read out of the frame store and the
timetag is applied.  This is further complicated by the spacecraft
dither which moves the image on the detector, causing the delay to
vary accordingly. We have written a simple ftools script that corrects
for this delay. The axBary program is then applied to correct the
times to the solar system barycenter.

We have folded the data on the period derived with our \rxte
observations of 5.17282 s (\cite{1056}).  We used a spline fit to the
data to define a pulse profile.  This profile is then cross-correlated
with 776 s subsets (i.e., 150 cycles each) of the data and the time
offset defines a local measurement of the phase (Figure 1). The upward
trend indicates that the best fit period for the \chandra data
(5.17293(4) s) differs slightly but significantly from the \rxte
value.  The period measurement errors are given in parenthesis and
represent the 1$\sigma$ uncertainty in the least significant digit. 
For the second \chandra observation, we find good agreement between
our measurement ($P = 5.17293(5)$ s) and the \rxte ephemeris.

We see in Figure 1 that there are systematic deviations
suggestive of a sine wave.  When we allow for a sine wave in our
fitting we obtain an F statistic of 7.0, which for the 3 additional
parameters is significant at about the 92\% level.  We do not feel
that these residuals reflect a real orbital period; they are most
likely due to phase noise (it is quite possible that the timing noise
detected here is responsible for the disagreement between the periods
derived with \rxte and \chandra). However, since the period (5500 sec
or about 92 min) would be extremely difficult for a satellite in low
Earth orbit to detect, we feel that it is important to set here an
upper limit; the 90\% confidence upper limit on the period amplitude
is 0.25 s. 

We compared the pulse profile in different energy bands ($0.5-1.7$,
$1.7-3.0$, and $3.0-7.0$ keV) to investigate shape-energy dependence.
We do not see any significant differences in the pulse shape with
energy, within statistics. We find that the 0.5$-$7 keV rms pulse
fraction is 16.2(9), and 14(1)\% during the first and second
observation, respectively.

\subsection{Spectral Analysis}

We extracted a six pixel wide spectrum centered at the peak of the
source image. Since in the continuous clocking (CC) mode, events
anywhere in a CCD column will be read out at the same collapsed CC
location, this region maps onto the sky as a rectangle
3$^{\prime\prime}$ high and 8.3$^{\prime}$ long rotated by the
spacecraft roll angle.  To allow normal $\chi^2$ fitting, we grouped
the spectrum into bins, which contain a minimum of 40 counts each.

We used CIAO (Version 2.1.1) to compute a response matrix assuming a
point
source at the center of the CCD.  It is also worth noting that the CC
mode used on-orbit was never calibrated on the ground.  Although the
on-orbit version of continuous clocking better approximates Time
Exposure (TE) mode data and therefore the CIAO computed matrix is a
good approximation, minor differences between the data and model
should be treated with a good deal of skepticism.

We used XSPEC (\cite{arnaud96}) to fit the spectra.  The measured
spectral parameters and $\chi^2$ values for all fits are listed in
Table 1.  For both data sets we find that the spectrum can be fit with
an absorbed power law; the corresponding unabsorbed ($2-10$ keV)
fluxes are 1.20(3) and 1.00(3) $\times10^{-11}$ ergs/cm$^2$ s, for the
first and the second observation, respectively. We find that in the
second data set the flux decreases by 20\% from the first observation,
in agreement with what has been reported by \cite{1055}, and that the
spectral index is slightly steeper (by 0.1).  

To optimize the statistics we added the two data sets; Table 1 shows
that a single power law model is a poor fit. To improve our fit we (i)
ignore channels between 1.8 and 2.5 keV since most of the residuals
occur from the Si K line (from the detector) to the Ir edge (from the
mirror) and hence are probably instrumental, and (ii) we add a
blackbody (BB). The BB addition gives an F statistic of 20.3; the
probability that the $\chi^2$ improvement is a chance coincidence is
$4\times10^{-5}$. We find that 19\% of the unabsorbed ($2-10$ keV)
flux comes from the blackbody component, i.e., $0.21(5)\times10^{-11}$
erg/cm$^2$s. The best fit parameters are in agreement with our
earlier work (\cite{woods99a,woods01a}).

Finally, we took the best fitting blackbody parameters from the summed
spectrum and added it to our model for the first and second data sets. 
We did not allow the BB temperature nor the normalization to vary and
refit for the power law parameters and $N_{\rm H}$. The PL $+$ BB fit
fluxes are identical to their single PL values. Although the errors
are relatively large, it is interesting that the entire difference
between the two data sets can be modeled with the change of a single
model parameter (i.e., the power law normalization).

Figure 2 shows the data of the first \chandra observation fit with a
two-component (PL $+$ BB) spectral model (solid line). We do not
observe any significant features in the spectrum; we have
investigated, however, the significance of the small excess at 3.15
keV. The addition of a line in the model for the first data set (at
3.15 keV) decreases the F statistic by 6.7 which is less than
3$\sigma$. The significance is further reduced if one considers that
the residuals are similar in amplitude to the Ir edge features near
2.1 keV, and that a 2$\sigma$ deviation is expected somewhere in the
spectrum from chance alone. Further the feature is not seen in the
second data set nor in the summed set.  Therefore we do not feel that
the line is real.  However, we can use it to set an upper limit for a
line detection.  The 90\% confidence upper limit on the line rate is
$2.0\times10^{-5}$ photons/cm$^2$ s, corresponding to an equivalent
width of 38 eV.

\subsection{Search for extended emission}

Given the large column to \sgr we {\it expect} a dust scattering halo
surrounding this X-ray source. To search for this as well as for the
presence of any hypothetical X-ray nebula, we investigated the
1-dimensional CC image.  We found no significant point sources other
than the SGR; for each pointing we generated two images within the
0.5$-$7 keV energy band.  The first image is simply the
time-integrated image.  We approximated the internal instrumental and
diffuse Galactic plane background on chip S3 by using data from the
other
back-illuminated CCD S1.  We measured the average rate on S1,
corrected for the absolute normalization (between 0.5$-$7 keV) at the
same focal plane temperature between the two chips S1 and S3
(\cite{markevitch01}), and then subtracted the S1 average (DC) level
from chip S3.  

Next, we generated a ``pulsed'' image for each observation. We
convolved the event list for S3 with the observed source pulse profile
normalized to a mean of zero to generate a background-subtracted
image.  That is, events recorded near pulse maximum received a
positive weight and those during pulse minimum, a negative weight. 
Applying this technique, we removed all emission components in our
image that do not vary in phase (i.e., everything except the pulsar
photons).  

We compared the time-integrated images and found no significant
difference
between them. To improve our statistics, we then averaged first the
time-integrated images and separately the pulsed images for the two
observations and folded the 1-D images about each centroid to
accumulate
``quasi-radial" profiles (these are not true radial profiles because the
CC
mode data were first collapsed in one dimension).  The time-integrated
and
pulsed profiles are shown in Figure 3 in addition to a MARX PSF model
(dotted
line) scaled to the normalization of each.  As expected, the pulsed
image is
entirely consistent with the model PSF.  However, due to the small pulse
fraction of this source, the pulsed image has a relatively poor
signal-to-noise
ratio, which is reflected in the rather large error bars. We find a
significant
excess above the model PSF for the time-integrated image further than
3$^{\prime \prime}$ away from the centroid. This excess contributes
$<0.5$ \%
to the total flux and can be attributed to a dust halo. We discuss these
results in the next section.

\section{Discussion}

We have observed in the past that SGR activity results in an increase
of the persistent X-ray flux; the flux then decays following a
power-law function with varying temporal indices for different
sources. For \sgr, the `nominal' (baseline) 2$-$10 keV flux is $\sim$1
$\times$ 10$^{-11}$ ergs cm$^{2}$ s$^{-1}$ and the temporal decay
index was found to be 0.6$-$0.7 during its previous activation in 1998
(\cite{woods01a,ibrahim01}).  The recent flaring activity of the
source has decayed with a similar index of $\sim0.6$ (\cite{1055}).
These results seem to indicate an intrinsic luminosity relaxation time
constant per SGR source, that appears to be independent of the
activity level. 

We have shown earlier that the overall increase of the source
intensity is entirely due to the power law component of the spectrum
(\cite{woods99a}). At the peak of activity, the spectrum is best fit
with a single power law model; the thermal component reemerges only
when the power-law flux reduces to its `nominal' level.  Our \chandra
results confirm the blackbody component in the spectrum of \sgr, which
is the {\it only} SGR that shows clear evidence of thermal emission
from the neutron star surface.  With regards to its X-ray spectrum,
therefore, \sgr seems to provide yet another link with the other class
of objects identified as magnetars, namely the Anomalous X-Ray Pulsars
(AXPs). There are so far 5 (maybe 6) AXPs, sources that are very
similar to SGRs, with the main exception that they have never been
observed to burst (\cite{mereghetti99}).

The radio flux (3$\sigma$) upper limit we derived with MERLIN
$\sim$1.6$-$2.1 days following this flare is both closer in time and
slightly above the earlier reported detection level for \sgr during its
1998 reactivation. VLA measurements at the same wavelength detected a
possible peak in the flux of $\sim0.3$ mJy (\cite{frail99}) $\sim$10
days after the huge flare that released $\sim10^{44}$ ergs in
$\gamma$-rays. Frail et al. (1999) estimate the minimum energy release
in particles necessary to create the detected synchrotron nebula to be
of the order of $7\times10^{37}$ ergs; the intermediate burst that
triggered the current activity released 25 times less energy in
$\gamma$-rays.  If we scale the particle energy accordingly, we derive
an upper limit of $4\times10^{35}$ ergs for any transient synchrotron
nebula associated with this burst. Our observations show that a prompt
and relatively high-intensity radio flare immediately following an
\sgr outburst is unlikely. Unfortunately, there are no MERLIN or other
radio data available for \sgr at $\sim$10 days after its recent
outburst.

The observed profile of the extended emission is consistent with
scattering of the X rays on the interstellar medium along the line of
sight from the source, given the large value of the observed $N_{\rm
H}$ (\cite{predehl95}). This is the first detection of a dust
scattering halo around a magnetar; unfortunately, due to the large
distance of \sgr, any pulsed component is smeared out in the halo
preventing the direct measurement of its distance using its X-ray
scattering halo (\cite{predehl00}). 

In conclusion, we have provided a constraining upper limit on the
radio flux from a transient synchrotron bubble following the
intermediate flare of 18 April 2001.  We confirmed the thermal
component in the spectrum of \sgr, and have placed a limit on line
features in the X-ray spectrum.  Utilizing the sensitivity and
resolving power of \chandra, we have discovered a dust scattering halo
surrounding \sgr.  We expect that our upcoming \chandra observation
(NRA2) will allow for a more detailed study of this halo.

{\bf Acknowledgements:} We acknowledge support from the following
grants: MX-0101 (C.K.), NAG5-9350 (C.K., P.W. and S.P.), GO0-1018X
(S.P.), JPL Contract 958056 (K.H.). We are grateful to Peter Thomasson
for rapidly arranging the MERLIN scheduling at a very short notice and
to Dr. C. Thompson, Dr. R. Duncan, Dr. P. Predehl and Dr. M.S. Briggs
for critical discussions on our results.


\clearpage

\begin{figure}[h]
\plotone{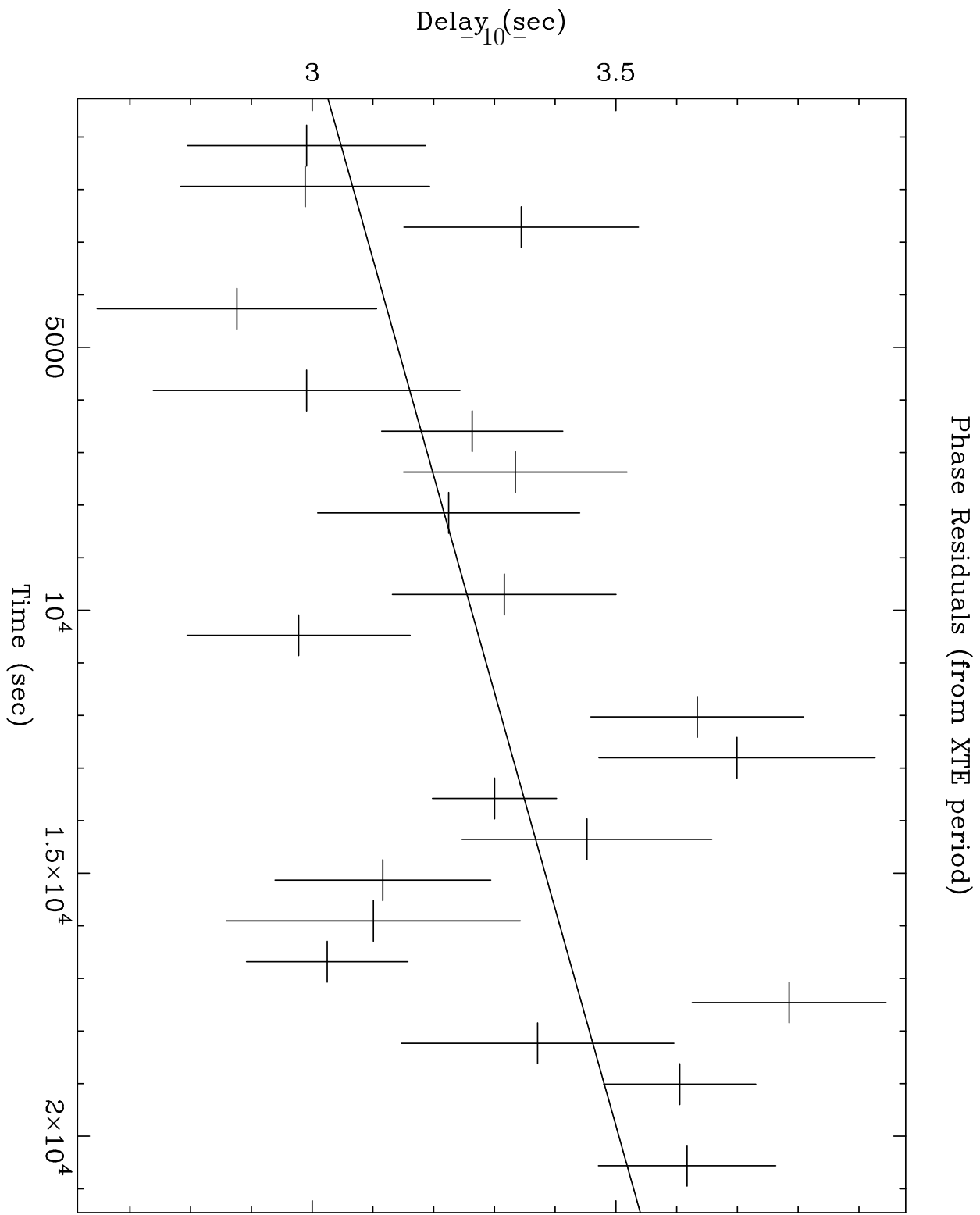}
\caption{The \chandra phase residuals (2$-$10 keV) from the \rxte PCA
ephemeris for the first observation on April 22.  The solid line is a
fit to the residuals with a straight line.}
\label{onebarrel}
\end{figure}                                                

\begin{figure}[h]
\plotone{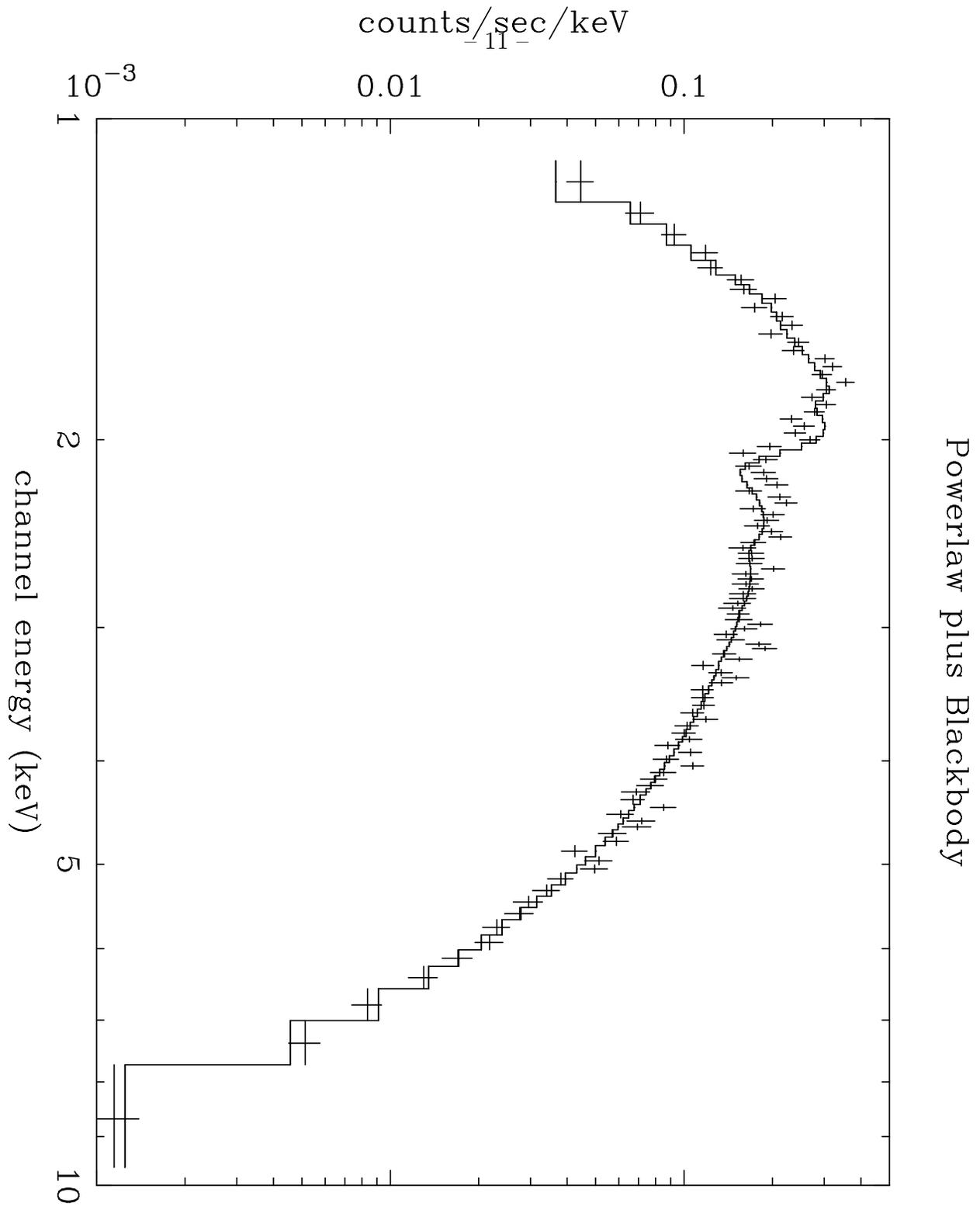}
\caption{The energy spectrum of the first \chandra observation of
\sgr fitted with a power law $+$ blackbody model (solid line).}
\label{onebarrel}
\end{figure}                                                

\begin{figure}[h]
\plotone{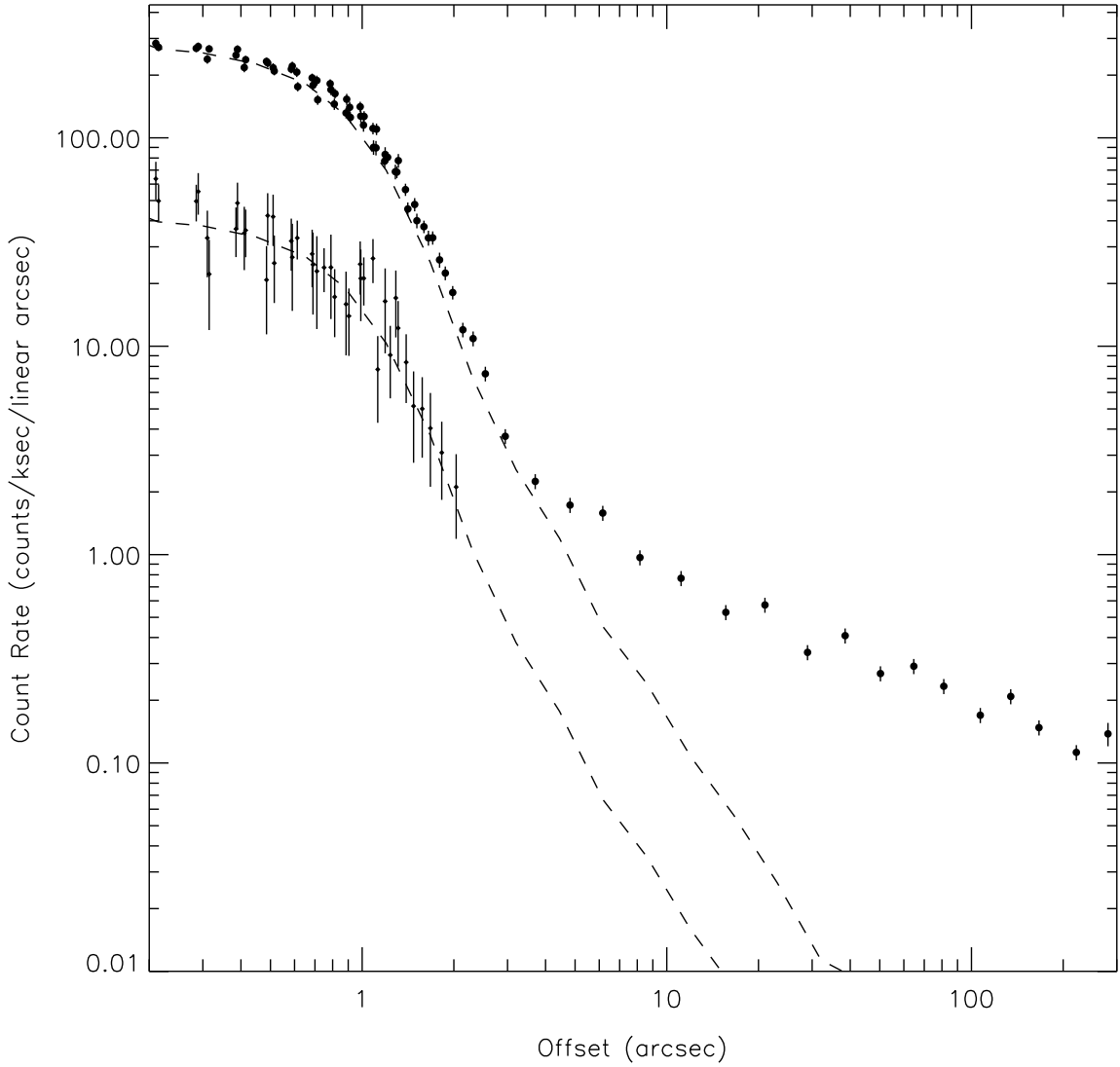}
\caption{The time-integrated (filled circles) and pulsed (diamonds)
profiles of \sgr averaged over both \chandra observations. The dotted
lines represent the PSF models scaled to the normalization of each
profile.}
\label{onebarrel}
\end{figure}                                                

\clearpage


\begin{deluxetable}{cccccc}
\tablecaption{Summary of spectral fits. \label{tbl-1}}
\tablewidth{0pt}

\tablehead{
\colhead{Date}  &  \colhead{Model\tablenotemark{a}}    &  
\colhead{N$_{\rm H}$}   &  
\colhead{$\Gamma$}          &
\colhead{$kT$}   &  \colhead{$\chi^2$/dof}  \\
\colhead{}  &  \colhead{}    &  
\colhead{(10$^{22}$~cm$^{-2}$)}   &  
\colhead{}          &
\colhead{(keV)}   &  \colhead{} 
}

\startdata

4/22     &  PL     &     2.69(10)    &  2.66(9)    &  
        ...            &     232.8/211          \nl

4/22     &  BB+PL  &     2.44(9)    &  2.07(10)     &  
        0.46           &     233.3/211              \nl

4/30     &  PL     &     2.77(12)    &  2.82(10)     &  
        ...            &     210.1/173           \nl

4/30     &  BB+PL  &     2.40(9)    &  2.07(13)     &  
        0.46           &     197.8/173             \nl

Both         &  PL     &     2.73(8)    &  2.73(7)     &  
        ...            &     367.7/272           \nl

Both         &  BB+PL  &     2.3(2)      &  2.1(3)       &  
   0.50(4)     &     349.7/279            \nl

Both\tablenotemark{b}         
             &  PL     &     2.71(8)    &  2.73(7)     &  
        ...            &     240.5/223    \nl

Both\tablenotemark{b}         
             &  BB+PL  &     2.4(2)      &  2.1(3)       &  
   0.46(5)     &     220.2/221          \nl

\enddata

\tablenotetext{a}{PL = power law; BB = blackbody}
\tablenotetext{b}{Energy channels between 1.8 and 2.5 keV ignored}

\end{deluxetable}
\end{document}